\documentclass[aps,prl,superscriptaddress,unsortedaddress,twocolumn,showpacs]{revtex4}

\usepackage{amsfonts}

\usepackage[pdftex]{graphicx}
\usepackage{epstopdf}
\usepackage{amsmath}
\usepackage{amssymb}
\usepackage{tikz}
\usepackage{comment}

\usepackage{ulem} 

\newcommand{\tialp }{\widetilde \alpha}
\newcommand{\tik }{\widetilde k}

\definecolor{linkcolor}{rgb}{0,0,0.6} 
\usepackage[pdftex,colorlinks=true, pdfstartview=FitV, linkcolor= linkcolor, citecolor= linkcolor, urlcolor= linkcolor, hyperindex=true,hyperfigures=true]{hyperref} 

\begin{document}  
 \title{Thermal bath Engineering for  Swift Equilibration}

\author{Marie Chupeau}
\affiliation{LPTMS, CNRS, Univ. Paris-Sud, Universit\'e Paris-Saclay, UMR 8626, 91405 Orsay,
  France}
\author{Benjamin Besga }
\affiliation{Universit\'e de Lyon, CNRS, Laboratoire de Physique de l'\'Ecole Normale Sup\'erieure, UMR5672, 46 All\'ee d'Italie, 69364 Lyon, France.}

\author{David Gu\'ery-Odelin}
\affiliation{Laboratoire de Collisions Agr\'egats R\'eactivit\'e, CNRS UMR 5589, IRSAMC, France }

\author{Emmanuel Trizac}
\affiliation{LPTMS, CNRS, Univ. Paris-Sud, Universit\'e Paris-Saclay, UMR 8626, 91405 Orsay,
  France}
  
\author{Artyom Petrosyan}
\affiliation{Universit\'e de Lyon, CNRS, Laboratoire de Physique de l'\'Ecole Normale Sup\'erieure, UMR5672, 46 All\'ee d'Italie, 69364 Lyon, France.}

\author{Sergio Ciliberto}\email[E-mail me at: ]{sergio.ciliberto@ens-lyon.fr}
\affiliation{Universit\'e de Lyon, CNRS, Laboratoire de Physique de l'\'Ecole Normale Sup\'erieure, UMR5672, 46 All\'ee d'Italie, 69364 Lyon, France.}

\begin{abstract}

We propose a new protocol that ensures the fast equilibration of an overdamped harmonic oscillator by a joint time-engineering of the confinement strength and of the effective temperature of the thermal bath. We demonstrate experimentally the effectiveness of our protocol with an optically trapped Brownian particle and report an equilibrium recovering time reduced by about two orders of magnitude compared to the natural relaxation time. Our scheme paves the way towards reservoir engineering in nano-systems.
\end{abstract}

\maketitle

Accelerating the relaxation dynamics of a system driven from an equilibrium state to a new one is an expedient goal and
a widely studied problem,  for its useful potential applications 
in quantum~\cite{PRX2014,Jarzynski_STA_2017,torrontegui2013,CKR08,SSV10,SSC11,BVM12,Bowler,Walther,Rohringer,Zhou,Du} or classical systems~\cite{GOMRMT14,ESE_2016,APL_AFM_2016,PRE_2017_Li} 
and for photonics circuit design~\cite{TsC12,Tse14,HTs14,Ste14}. Acceleration is achieved by engineering protocols 
which  shape  the dynamics of several control parameters to equilibrate  a system 
much faster than its characteristic relaxation time. Such an approach, known in the  engineering community  as input shaping~\cite{Singer,Devasia}, has been recently extended to  
microscopic systems where quantum and thermal fluctuations cannot be neglected~\cite{peercy2000}. For isolated quantum and classical systems, several techniques known as Shortcut to 
Adiabaticity (STA) have been developed and successfully applied to experiments~\cite{CKR08,SSV10,SSC11,BVM12,Bowler,Walther,GOMRMT14,ESE_2016,APL_AFM_2016,PRE_2017_Li,Rohringer,Zhou,Du}. 
In a recent article, we extended STA to systems coupled to a heat bath by introducing the so-called Engineered Swift Equilibration (ESE) processes~\cite{ESE_2016,APL_AFM_2016}. 
This extension is a key step for a number of applications in nano oscillators~\cite{kaka2005}, in the design of nanothermal engine~\cite{martinez2015,heat_engine_Lutz,heat_engine_Bechinger}, 
or in monitoring mesoscopic chemical or
biological processes~\cite{collin2005}, for which thermal fluctuations are of paramount importance and 
an accelerated equilibration desirable for improved power.
Specifically, we applied ESE to shorten by about 2 orders of magnitude the relaxation time of a micro-mechanical oscillator~\cite{APL_AFM_2016} and of a Brownian particle 
trapped in a harmonic potential whose stiffness has been suddenly increased (compression of the position distribution of the particle)~\cite{ESE_2016}. 

In this letter, we propose and test an original protocol based on a random forcing that allows
us to overcome an important experimental limitation for protocols in which the potential stiffness is reduced, (expansion of the trapping volume
leading to a decompression of the position distribution of the particle). For sufficiently fast transformations, both STA and ESE protocols generate solutions in which 
the external confinement becomes transiently repulsive, {\it i.e.} with a negative spring constant~\cite{torrontegui2013,GOMRMT14}. 
Indeed, the reverse curvature yields an exponential acceleration of the transformation.

However the experimental implementation of such a protocol is not always feasible or may be so cumbersome to implement that it is in practice unrealistic. 
This is the case for mechanical oscillators and colloidal Brownian particles. A second adverse feature of decompressions as opposed to compressions is that they are associated to an increasing intrinsic relaxation
scale, so that the adiabatic route becomes more time consuming.

In this letter, we show both theoretically and experimentally that a protocol based on a thermal bath engineering combined with a proper control of the strength of 
the confinement outperforms  STA and ESE protocols, bypassing the requirement for a transient repulsive potential. The key ingredient to allow for thermal bath time-control is 
to monitor the amplitude of a white noise random forcing which impinges on the system through an effective temperature. 
The corresponding original Thermally Engineered Swift Equilibration (TESE) protocol is certainly useful in a wealth of other applications where a fast expansion is needed.

We consider an overdamped system described by a first order Langevin equation for the coordinate $x$ of a Brownian particle trapped in a harmonic potential of stiffness $k(t)$ centered in $x_o(t)$:
\begin{equation}
\nu \dot x= -k(t) \, (x-x_o(t)) + \sqrt{2\, k_BT  \, \nu} \,  \xi(t) 
\label{eq:langevin}
\end{equation}
where $k_B$ is the Boltzmann constant, $T$ the temperature, $\nu$ the viscous coefficient and $ \xi$ a delta correlated noise such that $\langle \xi(0)  \xi(t)\rangle=\delta(t)$. 
In equilibrium when $x_o(t)=0$ and $k$ is constant, the characteristic relaxation time is $\tau=\nu/k$, and the probability density function $P(x)$ of the position fluctuations is Gaussian with standard deviation $\sigma= \sqrt{k_BT/k}$. 

We imagine a process where $k$ is changed from an initial  constant value $k_i$ to a final value $k_f$. This process can be done by a STEP protocol in which  $k$ 
is suddenly changed  from $k_i$ to $k_f$ and   the system relaxes to the new equilibrium state in several $\tau$, which is the time taken by $\sigma$ 
to reach the final value $\sigma_f$. More interestingly, the ESE protocol in which $x_o=0$ allows us to arbitrarily reduce the relaxation regime using an 
appropriate time evolution for $k$.
Let us briefly recall how this protocol is constructed.
We first consider  the dynamics of $P(x,t)$, which is described by the Fokker-Planck equation  associated to Eq.~\eqref{eq:langevin}:
\begin{equation}
 \partial_t P \, = \, {k(t)\over \nu} \, \partial_x(x \, P) \, + \, {k_BT\over \nu} \, \partial^2_{x} \,P.
\label{eq:FP}
\end{equation}
As  during the time evolution $P(x,t)$ remains Gaussian, we write $P(x,t)= (\alpha(t)/\pi)^{1/2} \exp{(-\alpha(t)\, x^2)}$, where $\alpha$ is related 
to the variance of $x$ through $\alpha(t)=1/[2\sigma(t)^2]$. Thus Eq.~\eqref{eq:FP} becomes 
\begin{equation}
{\dot{\alpha} \over \alpha}= {2 k(t)  \over \nu}-{4 k_BT \alpha \over \nu}
\label{eq:alpha}
\end{equation}
where the dot denotes the temporal derivative, and the boundary conditions at initial time $t=0$ and final time $t=t_f$ are 
$\alpha(0 )  =   k_i / 2 k_BT$, $\alpha(t_f)=k_f / 2 k_BT$, and $\dot \alpha(0 )= \dot \alpha(t_f )=0$.
The conditions on the derivatives are important because they ensure equilibrium for $t\ge t_f$. 
The problem can be solved, using dimensionless quantities $\tialp=\alpha/\alpha_i$, $\tik=k/k_i$, $\chi=k_f/k_i$ and $s=t/t_f$, choosing a specific time evolution for $\tialp$. We subsequently determine the time dependence of $\tik$ which allows the system to equilibrate in an arbitrary time $t_f$: 
\begin{equation}
 \tik (t) \,=\,  \left(1+ {\chi  \,  \tau_f \,\partial_s \tialp \over 2 \, t_f \,\tialp^2 } \right) \tialp
\label{eq:tik}
\end{equation}
where $\tau_f=\nu /k_f$. A reasonable choice  for $\tialp$, which fulfills the boundary conditions,  is for example:
\begin{eqnarray}
\tialp(s) & =  & 1 {\rm \, \,  \,   for \, \,  \,}  s<0  \notag \\
  \tialp (s)& = &1+(\chi-1) (3 s^2-2 s^3) {\rm \, \,  \,   for \, \,  \, } 0\le s\le1 \notag \\
\tialp (s) & =  &\chi     {\rm \, \,  \,   for \, \,  \, }  s>1 
\label{eq:tialpsol}
\end{eqnarray}
which can be applied in a compression $\chi>1$ and in an expansion $\chi<1$.  In Ref.~\cite{ESE_2016}, we have shown that in the compressive case, 
$\tik$ following from Eqs.~\eqref{eq:tik} and \eqref{eq:tialpsol} can be successfully used in an experiment with a $t_f$ 
two orders of magnitude smaller than the natural equilibrium recovery time of several $\tau_f$. Yet, it is straightforward to see that for a quick expansion 
($t_f \ll \tau_f$), the stiffness may take negative values; it necessarily does so when $t_f$ is small enough, irrespective of the 
specific form chosen in Eq. (\ref{eq:tialpsol}). 
Thus, the potential has to be transiently repulsive in order to reach equilibrium within an arbitrary small $t_f$ \cite{GOMRMT14}. This constitutes a substantial experimental shortcoming since it is in general arduous if not impossible to turn a trapping (attracting) potential into a repulsive one, 
following the complex dynamical rule embedded in Eq. (\ref{eq:tik}).

\begin{figure}[ht!]
\centering
\includegraphics[width=0.98\columnwidth]{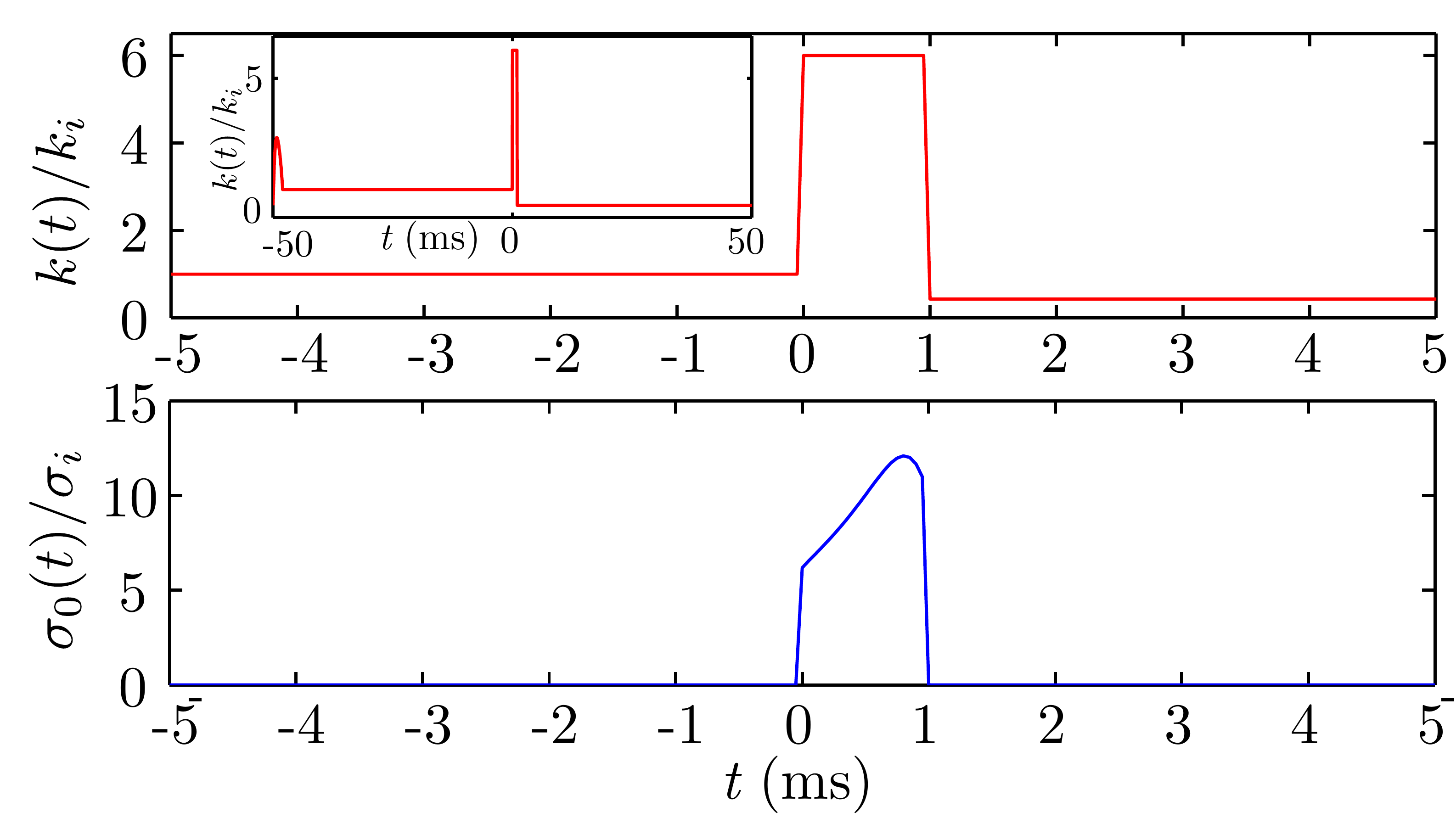}
\caption{The two  signals $\tik (t)$ and $\sigma_o(t)$ used to drive fast expansion. 
The time is set to zero when the decompression is triggered. The inset exhibits the full dynamics used in the experiment, cycled $4. \,10^4$ times. 
 The system is initialized at $t=-50\,$ms with an ESE-compression from $\tik_f$ to $\tik_i$, (noiseless TESE with Eqs.~\eqref{eq:tik} and \eqref{eq:tialpsol}),
 then left unperturbed, before the decompressing-TESE is applied at $t=0$. The system is subsequently left unperturbed up to $t=50\,$ms, before the cycle repeats. 
 The TESE sector ($0\leq t\leq t_f=1\,$ms)
 is ruled by Eqs.~\eqref{eq:tialpsol}, \eqref{eq:tik_evo} and \eqref{eq:noise}.
 Here, $\tau_f=11$ ms, $\tau_i/\tau_r=100$, $\chi\equiv k_f/k_i=0.44$ and $\tik_{\rm max}=6$.
}
\label{fig:ESE_driving}
\end{figure}

However, this problem can be resolved by a careful monitoring of the extra control parameter $x_o(t)$ (trap center's position), that vanished  in the previous analysis. 
We choose for  $x_o(t)$ an externally  generated random noise with correlation time 
$\tau_c \ll \tau_i$, {\it i.e.} the noise can be considered delta-correlated for all practical purposes with a time-dependent amplitude $\sigma_o(t)^2=\langle x_o^2 \rangle$. 
The correlation is chosen  as $\langle x_o(t) x_o(t') \rangle \,= \,\sigma_o(t)^2 \,\tau_r \, \delta(t-t')$,
where $\tau_r$ defines the noise spectral density.  This noise consequently behaves as another heat bath for the Brownian particle, 
with an effective temperature $T^*(t)=k(t)^2 \sigma_o(t)^2 \tau_r/(2 \, \nu k_B)$, so that
\begin{equation}
 \partial_t P = {k(t)\over \nu} \partial_x(x \, P) + \left({k_BT\over \nu} +{ k(t)^2 \sigma_o(t)^2 \tau_r \over 2 \nu^2}\right) \partial^2_{x} P 
\end{equation}
and the equation for $\tialp$ becomes
\begin{equation}
{ \chi \, \tau_f \, \partial_s \tialp \over t_f \tialp}= {2 \tik(t) }-{2 \tialp }- {{\tik(s)^2 \sigma_o(s)^2  \tialp (s) \tau_r \over  \sigma_i^2 \tau_i}}.
\label{eq:alpha_noise}
\end{equation}
We keep for $\tialp$ the same time evolution as in Eq.~\eqref{eq:tialpsol} but for the sake of convenience
 we fix $\tik(s)$ as 
\begin{eqnarray}
 \tik(s) & =  & 1 {\rm \, \,  \,   for \, \,  \, } s<0  \notag \\
  \tik (s)& = &  \tik_{\rm max}   {\rm \, \,  \,   for \, \,  \, } 0\le s\le1 \notag\\
 \tik (s) & =  &\chi     {\rm \, \,  \,   for \, \,  \, }  s>1 \label{eq:tik_evo}
\end{eqnarray}
where $\tik_{\rm max}>1$  is a chosen fixed value. It is interesting to note that in bypassing ESE that 
requires negative $k$-values to reach $k_f < k_i$, we make transient use of $k>k_i$. We take advantage of this feature, see below.
Having specified the dynamics of $\tialp$ and $\tik$, we can solve Eq.~\eqref{eq:alpha_noise} for the dynamics of $\sigma_o(s)$ 
\begin{equation}
\sigma_o(s)^2=\left[{1\over \widetilde \alpha \widetilde k}-\left(1+{\chi \, \tau_f  \,  
\partial_s   {\widetilde \alpha}\over 2 t_f \widetilde \alpha^2 }\right) {1\over \widetilde k^2} \right] {2 \, \sigma_i^2 \tau_i \over \tau_r} .
\label{eq:noise}
\end{equation}
In doing so, we have two experimentally controlled  parameters ($\sigma_o$ and $k$, see Fig.~\ref{fig:ESE_driving}), which allow us to 
follow the proper dynamics for $\alpha$ and to reach equilibrium exactly at $t_f$.

From Eq.~\eqref{eq:noise}, we notice that the value of $\tik_{\rm max}$
determines the maximum amplitude of  the noise, which decreases  when $\tik_{\rm max}$ increases.
Of course, one can make other functional choices for $k(t)$ and $\alpha(t)$. One may for instance wish to minimize the work performed during the transformation, 
but for experimental reasons and limitations, the chosen simple scheme seems the most appropriate.

We have implemented the TESE expansion protocol on a Brownian particle trapped by an  optical tweezer \cite{Berstat16}. 
Our experimental system  consists of a silica microsphere of radius $R=1\;\mu$m ($\pm 5\%$) immersed in water. The particle is trapped by an optical harmonic potential 
$U(x,t)=k(t)(x-x_o(t))^2/2$, where $x$  is the particle position and $x_o(t)$ is the focal position of the beam.   The stiffness  $k(t)$ of the potential  can be controlled by the power of the trapping laser. 
The fluid chamber is specifically designed in order to have only one bead in the measuring region,  which allows us to perform very long measurements without any perturbation induced by the other particles. The trap is realized using  a near infrared laser beam ($\lambda=1060$\,nm with maximum power 500\,mW) expanded and inserted through an 
oil-immersed objective (Leica, 63$\times$ NA 1.40) into the fluid chamber. The trapping laser power
is modulated by an external voltage $V_k$ via  a laser diode controller 
with a rising time of about  $40\, \mu$s. $V_k$ is generated by a National Instrument card (NI PXIe-6663) managed by a custom made Labview program.  The detection of the particle position is achieved using 
an additional  laser beam (at $\lambda=635$ nm power 0.5 mW), which  is expanded and collimated by a telescope and passed through the trapping objective. The forward-scattered detection beam is 
collected by a condensor (Leica, NA 0.53), and its back focal-plane field distribution projected onto a custom Position Sensitive  Detector  
(band pass of 1MHz) whose signal is acquired at a sampling  rate of 50 kHz with a NI PXIe-4492 acquisition board. The trapping beam goes through an acousto-optic deflector (AOD) that allows us to control the central  position $x_o(t)$ 
of the trap rapidly (up to $25$ KHz). 
In order to move the trap randomly, a Gaussian white noise voltage is generated by the analog output of a NI PXIe-6366 card and sent to the AOD. The conversion factor for the displacement due to the AOD is $A=5$ $\mu$m/V. 
If the amplitude of the displacement of the trap is sufficiently small to stay in the linear regime ($\sigma_o(t) <R$), it creates a random force 
on the particle that does not affect the stiffness of the trap. When the random force is switched on, the bead quickly reaches a stationary state 
with an effective temperature 
\cite{Berut_PRL,ber14}.

The stochastic command of the AOD is created with a Labview program that generates at a rate of $f_n=20$ kHz a white noise $x_d(t)$ of variance  $\sigma_d^2(t)$ and 
power spectral density $\sigma_d^2(t)/f_n$. It is then numerically low-pass filtered at frequency $f_{\rm lp}$ to produce $x_o(t)$, whose correlation time is $\tau_{\rm lp}=1/(2\pi\, f_{\rm lp}) \ll t_f<\tau_i$. The TESE protocol is not influenced by this filter provided that $f_{\rm lp} \gg k_{\rm max}/(2\pi \nu)$. In this regime, $x_o$ can be approximated as a white noise
\begin{equation}
\langle x_o(t) x_o(t') \rangle \simeq \frac{\left(A \,\sigma_d(t)\right)^2 }{2 f_n {\tau_{\rm lp}}} e^{-\frac{|t-t'|}{\tau_{\rm lp}}} \simeq \frac{\left(A\, \sigma_d(t)\right)^2}{f_n} \delta(t-t'),
\label{eq:sigma2_xd}
\end{equation}
and choosing $\tau_r=1/f_n$, one gets $\sigma_d(t)=\sigma_o(t)/A$. This equation, together with Eq.~\eqref{eq:noise}, allows us to compute the amplitude $\sigma_d(t)$ of the driving of the noise needed to carry out a fast expansion. We emphasize that the choice of a constant $k(s)$ in Eq.~\eqref{eq:tik_evo} with a large $k_{\rm max}$ is useful to easily fulfill the condition $\sigma_o<R$, so as to remain in the harmonic approximation of the confinement.

\begin{figure}[ht!]
\centering
\includegraphics[height=0.45\columnwidth]{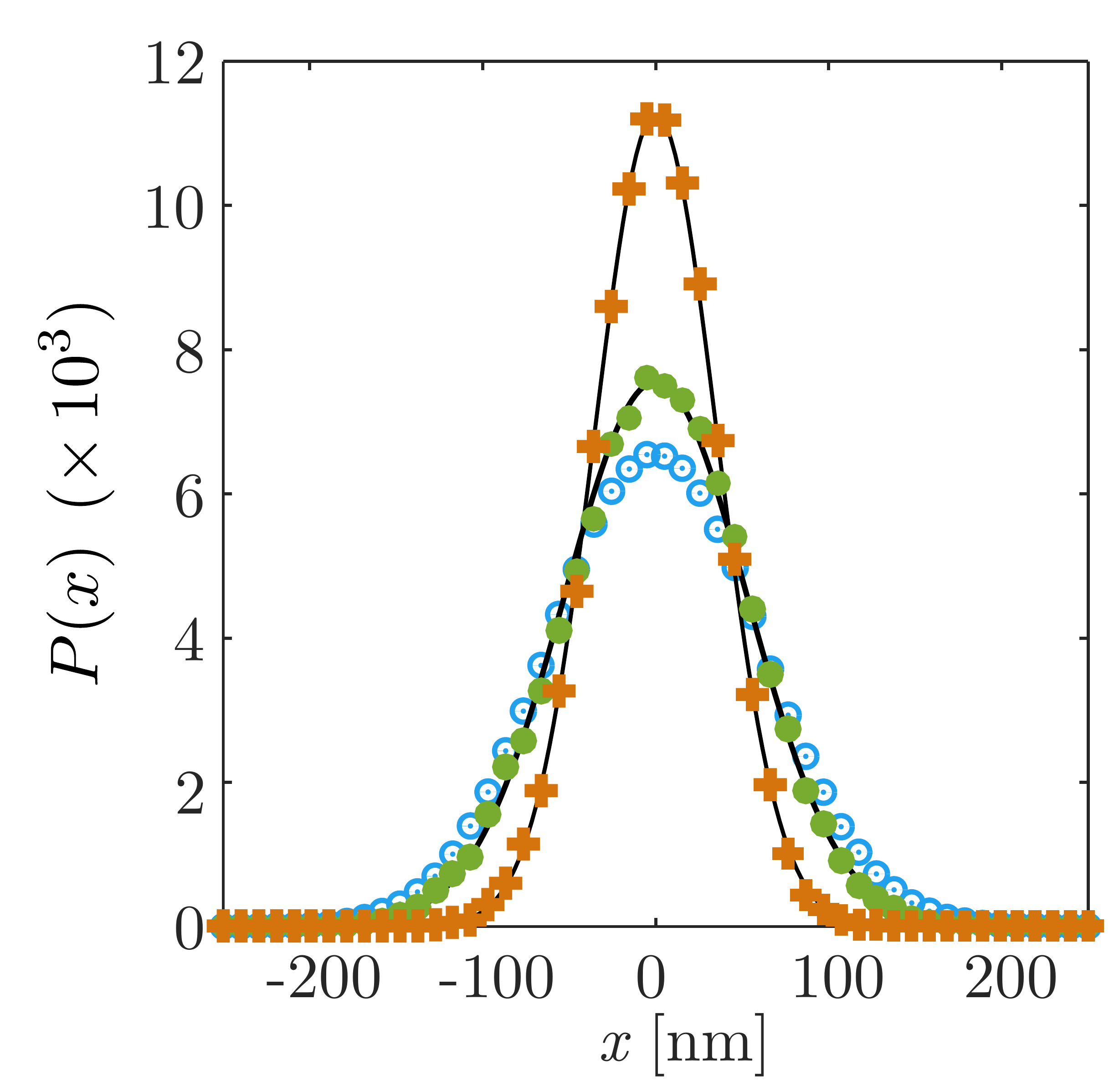}
\includegraphics[height=0.45\columnwidth]{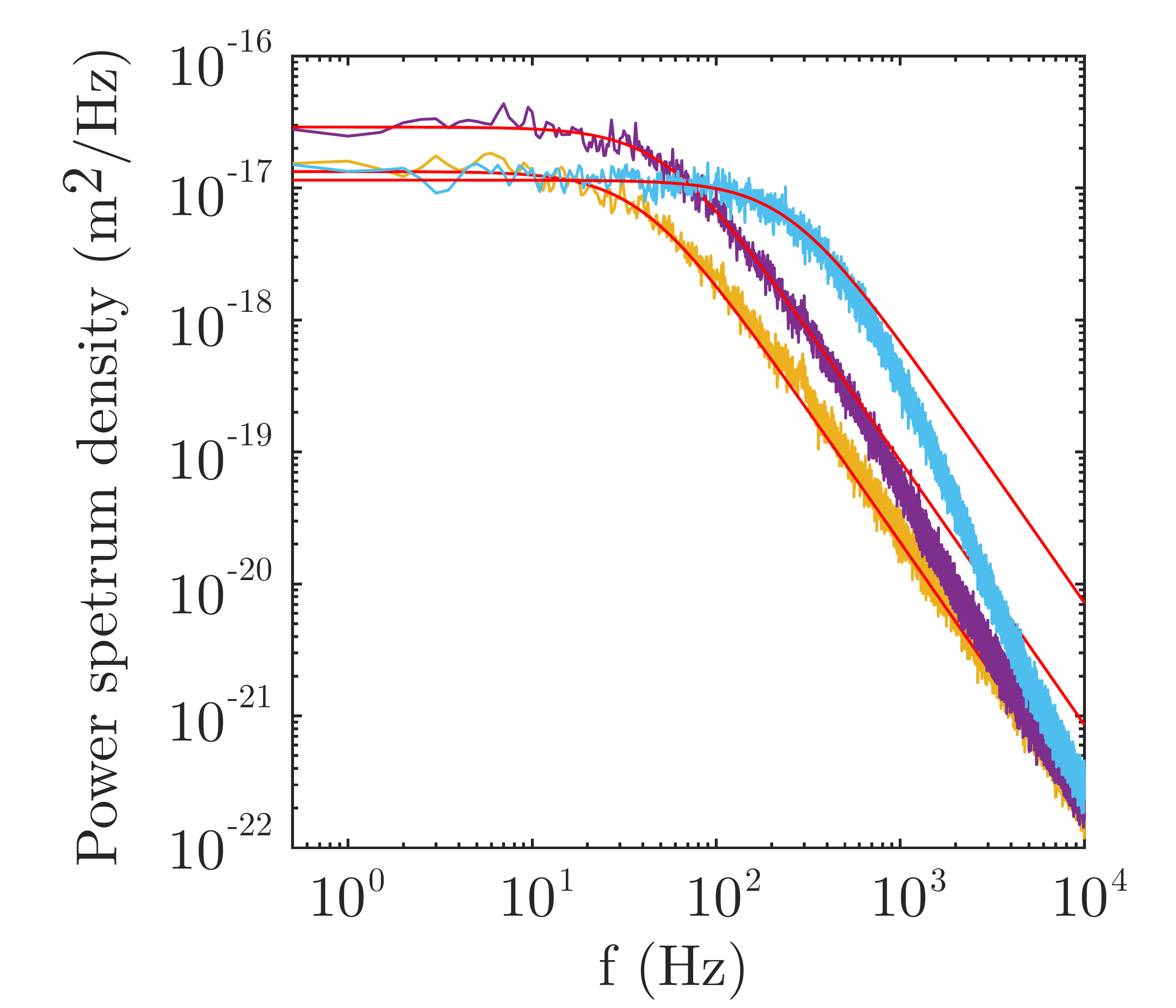}
\caption{(a) Density $P(x,t)$ measured in the initial compressed state (orange crosses), and at the end of the protocol at $t=t_f$ (green dots). 
The measured pdf are in  excellent agreement with the target densities shown with lines, for a decompression factor $\chi = 0.44$. 
We also measure the stationary pdf when the random force is switched on (blue circles, for a higher stiffness $k=4k_i$). (b) Power spectra of $x$ 
measured in equilibrium at $k=k_i$ without (orange line) and with noise (purple line). In this case, $\sigma_i=30$ nm, $f_c=41$ Hz and $\sigma_o=20$ nm. 
We also show a power spectrum for $k=4k_i$ with noise ($\sigma_o=21$ nm, blue line) corresponding to the blue circles in (a).
The Lorentzian fits allow us to calibrate the system and the noise amplitude. Beyond the filtering frequency $f_{\rm lp}=1$ kHz, 
the fits become meaningless and all curves eventually collapse. }
\label{fig:PSD}
\end{figure}

The system is calibrated in equilibrium using standard methods. The trap stiffness is  measured by calculating the variance of the $x$-displacement of the bead, $\sigma^{2}$. The other parameters are determined using the power spectrum of the $x$-displacement which is a Lorentzian since the particle motion is  overdamped: $S_k(f) = \frac{4 \nu k_{\mathrm{B}}T/k^2}{1+f^2/f_c^2}$. One can fit it to find the cut-off frequency $f_c$ that verifies $f_c = \frac{k}{2\pi {\nu}}$ where $\nu = 6\pi R \eta$ and $\eta$ is the dynamic viscosity of water. The two methods give compatible results for the stiffness as  the viscosity of water and corrections due to the finite distance between the particle and the bottom of the cell are known. This comparison allows us  to check the calibration of the bead position measurement. 

The probability density function (pdf) of the bead position is shown in Fig.~\ref{fig:PSD}(a) at the initial and final time of the protocol, displaying the expected decompression from $\sigma_i = 30$ nm to $\sigma_f= 45$ nm. Adding noise on the center of the trap broadens drastically the distribution, even if the stiffness is increased simultaneously as shown by the stationary pdf displaying a $\sigma_{\mathrm stat} = 60 $ nm at $k=4k_i$.
The power spectra of the bead displacement in the $x$-direction without noise  and with noise 
are shown in Fig.~\ref{fig:PSD}(b), measured in equilibrium  at $k=k_i$. The displacement in the $y$-direction is not modified by the added noise. Thus the noise behaves correctly 
and the amplitude is  the one computed from Eq.~\eqref{eq:noise} with $\dot \tialp=0$ and Eq.~\eqref{eq:sigma2_xd}. 
\begin{figure}[ht!]
\centering
\includegraphics[width=0.8\columnwidth]{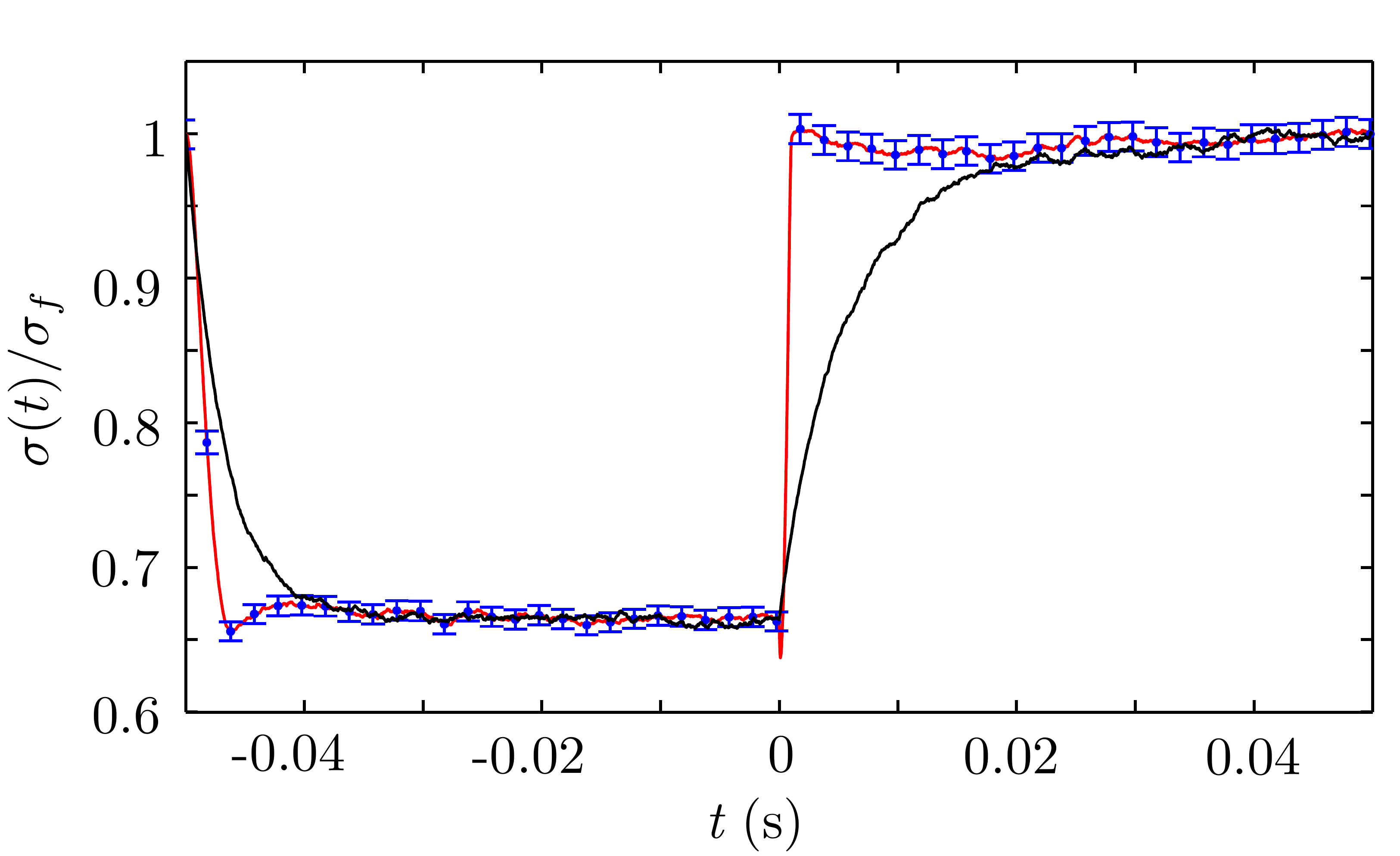} 
\caption{Dynamics of the standard deviation $\sigma(t)$ of the Brownian particle position, for two distinct cyclic schemes. The first one alternates between ESE for compression phases and TESE for decompression phases (red line), while the second one is only constituted of STEP procedures (black line). Time $t=-50$ ms corresponds to the beginning of the compression phase, where the stiffness is changed from $k_f$ to $k_i$, and time $t=0$ is the beginning of the expansion phase, where $k$ is switched from $k_i$ to $k_f<k_i$
(see also Fig. \ref{fig:ESE_driving}). The experimental measures are shown, together with error bars. 
To gather relevant statistics, the scheme is repeated as explained in the main text and in 
Fig. \ref{fig:ESE_driving}. }
\label{fig:ESE_Square}
\end{figure}

In order to put the TESE protocol to the test, we prepare the system in an initial state with $k_i=3.3 \,  \rm{pN}/\mu\rm{m}$ having $\, \tau_i=4.7\,$ms and we commute to a 
final state $k_f=1.4 \,  \rm{pN}/\mu\rm{m}$ having $\tau_f=11\,$ms.  After $50$ ms the system is commuted back to the initial state. This protocol is repeated $2.\, 10^4$ times 
to measure statistical quantities, among which the time evolution of $\sigma (t)$. As a benchmark of an ``uncontrolled'' process,
we first perform the commutation using  a STEP protocol in which the standard deviation of 
the particle position $\sigma (t)$ relaxes in a time that defines the natural recovery scale for either the initial or the final state. 
Then, the commutation is  performed using the ESE/TESE protocols presented in Fig.~\ref{fig:ESE_driving}. 
Specifically, the TESE protocol is applied to expand the trap from $k_i$ to $k_f<k_i$ whereas the standard ESE (Eqs.~\eqref{eq:tik} and \eqref{eq:tialpsol} without noise),
is used for the sake of cyclization, to compress the trap 
from $k_f$ to $k_i$ and thereby prepare the system for a new expansion. 
For the expansion, we chose $t_f=1\,$ms, thus significantly below the natural relaxation time $\tau_f$. We took $k_{\rm max}=6 \, k_i$,
which gives a maximum amplitude of $\sigma_o(t)$ of approximately 400\,nm, while for compression, $t_f=2$ ms.
The results are plotted in Fig.~\ref{fig:ESE_Square} where the STEP and TESE protocols are compared. We see that 
the STEP scheme requires 40 to 50 \,ms for equilibrium recovery, whereas equilibration is realized in $t_f \ll 50\,$ms 
within ESE (for compression) and TESE (for expansion).
Hence we have demonstrated that TESE drives the system to its new equilibrium about 2 orders of magnitude faster than the standard relaxation. This is the main result of this letter.
It is instructive here to comment briefly on the features of the bare ESE method that could in theory be used to
achieve a similar decompression. As emphasized above, it follows from Eqs.~\eqref{eq:tik} and~\eqref{eq:tialpsol} that the stiffness $k$ should take negative values for more than
80\% of the time span $[0,t_f]$, with a (negative) maximal peak amplitude $k_{\text{peak}} \simeq -2 k_i$. Devising 
a confinement system able to meet this goal, and furthermore follow the target ESE-dynamics, is a delicate experimental challenge.

To conclude, we have proposed a protocol, dubbed Thermally Engineered Swift Equilibration, that
takes advantage of a carefully shaped noisy signal, coupled to a harmonic time-dependent confining force,  to manipulate confined Brownian particles. It allows us to quickly deconfine 
a trapped particle, without the drawback  of passing transiently through a repulsive potential.  Related strategies have proven efficient in a computational
context, like simulated annealing \cite{Frenkel} and arguably, we provide here a transposition to
an experimental context.
The  protocol put forward   has been devised so as to remain very simple, with only two driving parameters $k_{\rm max}$ and $\sigma_o(t)$ that can be easily calibrated with an excellent accuracy. Interesting venues for future work
deal with optimizing the scheme. For a given time $t_f$, how can one minimize the energy needed to perform the TESE protocol among the infinitely many 
choices for coupled noise-confinement driving? This question is of importance in energetics, and a subject of intense 
current activity \cite{cui2015,ScSe07,ScSe08,Aurell,Deff15,Pol15,Muratore}. Besides, 
although we used here a harmonic potential, the TESE protocol can  be extended to non-harmonic cases, along the lines proposed 
in Ref.~\cite{ESE_2016}. This original protocol bears promises for applications whenever a controlled fast  expansion is needed, and 
the experimental constraints do not allow the generation of an expulsive potential. 
\section{Acknowledgments}
This work has been supported by the ERC contract OUTEFLUCOP. We acknowledge funding from the
Investissement d'Avenir LabEx PALM program (Grant No.
ANR-10-LABX-0039-PALM).
\vfill
\bibliographystyle{unsrtnat}
\bibliography{references_ese_V2.bib}

\end{document}